\documentclass[journal]{vgtc}                     
\usepackage{graphicx}    
\usepackage{microtype}                 
\usepackage{xcolor}
\usepackage[normalem]{ulem}
\newcommand{\amend}[1]{#1}
\newcommand{\remove}[1]{}

\onlineid{1468}

\vgtccategory{Research}

\title{From Vision to Touch: Bridging Visual and Tactile Principles for Accessible Data Representation}

\author{%
  \authororcid{Kim Marriott}{0000-0002-9813-0377},
  \authororcid{Matthew Butler}{0000-0002-7950-5495},
  \authororcid{Leona Holloway}{0000-0001-9200-5164},
  William Jolley,
  \authororcid{Bongshin Lee}{0000-0002-4217-627X},\\
  Bruce Maguire, and
  \authororcid{Danielle Albers Szafir}{0000-0003-1848-7884}
}

\authorfooter{
  \item
  	Kim Marriott, Matthew Butler, and Leona Holloway are with Monash University.
  	E-mail: \{kim.marriott, matthew.butler, leona.holloway\}@monash.edu
  \item
  	William Jolley is an independent researcher.
    Email: wjolley@bigpond.com
  \item 
  Bongshin Lee is with Yonsei University.
  E-mail: b.lee@yonsei.ac.kr
  \item 
  Bruce Maguire is with Vision Australia.
  E-mail: bruce.maguire@visionaustralia.org
   \item 
  Danielle Albers Szafir is with University of North Carolina Chapel Hill.
  E-mail: dnszafir@cs.unc.edu
}

\abstract{%
  Tactile graphics are widely used to present maps and statistical diagrams to blind and low vision (BLV) people, with accessibility guidelines recommending their use for graphics where spatial relationships are important. Their use is expected to grow with the advent of commodity refreshable tactile displays. However, in stark contrast to visual information graphics, we lack a clear understanding of the benefits that well-designed tactile information graphics offer over \remove{data tables or} text descriptions for BLV people. To address this gap, we introduce a framework considering the three components of encoding, perception and cognition to examine the known benefits for visual information graphics and explore their applicability to tactile information graphics. This work establishes a preliminary theoretical foundation for the tactile-first design of information graphics and identifies future research avenues. 
}

\keywords{Tactile graphic, visual perception, haptic perception, accessible data representation}

\teaser{
  \centering
  \includegraphics[width=0.98\linewidth, alt={}]{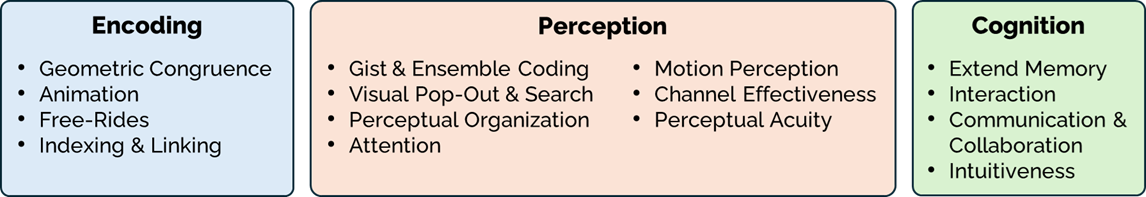}
  \caption{%
  	Our three component framework for analyzing the differences and similarities between visual and tactile information visualization. \amend{The factors in each component are drawn from Munzner~\cite{munzner2014visualization}, Streeb et al.~\cite{streeb2019visualize}, and Ware~\cite{ware2021information}, as well as our prior research\amend{~\cite{bae2025bridging,goncu2010tactile}}}.
  }
  \label{fig:teaser}
}

\renewcommand{\manuscriptnotetxt}{}

\graphicspath{{figs/}{figures/}{pictures/}{images/}{./}} 

\usepackage{tabu}                      
\usepackage{booktabs}                  
\usepackage{lipsum}                    
\usepackage{mwe}                       

\usepackage{mathptmx}                  

\newcommand{\bstart}[1]{\smallskip\noindent{\textbf{#1}}}

\begin{document}


\firstsection{Introduction}

\maketitle

We are witnessing a surge of research on how to design and support accessible data visualization and experiences, with a particular focus on blind or low vision (BLV) people~\cite{kim2021accessible,marriott2021inclusive,lee2024inclusive}. Researchers and practitioners have investigated various modalities to provide accessible data representation to BLV people, including text and audio-based approaches~\cite{siu2022supporting,thompson2023chart}, tactile and haptic-based approaches~\cite{Palani2020,Tennison2024}, and other combinations~\cite{seo2024maidr,zhang2024chart}.

Accessibility guidelines recommend the use of raised line drawings known as tactile graphics to present graphics that are inherently geometric in nature, such as maps and charts~\cite{BANA2022guidelines,RoundTable2022}. More recently, refreshable tactile displays (RTDs) enable the rapid generation of tactile graphics and dynamic interaction~\cite{reinders2024refreshable,Schwarz2022}. 

While studies have demonstrated the benefits of particular kinds of tactile graphics, such as maps~\cite{Papadopoulos2018}, scatterplots~\cite{WatanabeMizukami2018}, and node-link diagrams~\cite{Yang2020}, we currently lack a comprehensive understanding of the potential benefits of  tactile information graphics compared to alternative accessible representations \amend{such as} text descriptions. This gap stands in stark contrast to visual graphics, which have been extensively studied. Over the last four decades, researchers in visualization, philosophy, cognitive science, vision science, and psychology have investigated why visual information graphics are effective~\cite{streeb2019visualize}. Their findings range from mathematical arguments based on the nature of graphical representation, such as free-rides~\cite{shimojima2015semantic} and geometric congruence~\cite{larkin1987diagram}, to perceptual reasons like pattern recognition based on Gestalt principles~\cite{ware2021information} and cognitive benefits such as providing external memory~\cite{ware2021information} and support for collaboration~\cite[pp. 44-48]{hutchins1995cognition}. Similar theoretical foundations for tactile information graphics are largely unexplored.

To address this gap, we examine known benefits for visual information graphics and consider to what extent they also hold for tactile information graphics. We introduce a framework for this analysis (Section~\ref{sec:approach}) based on considering factors related to three components--encoding, perception, and cognition (Figure~\ref{fig:teaser}). In Sections~\ref{sec:encoding}, \ref{sec:perception}, and \ref{sec:cognition}, we examine the potential benefits and the similarities/differences in each of these components between visual and tactile information graphics. 
For example, touch readers cannot immediately see the overall shape or trends of a graphic and must build this structural knowledge through the sequential exploration of the graphic. 
This analysis provides a theoretical basis for further investigation into tactile information visualization and the design of effective tactile representations. The main contributions of this paper are:
\begin{itemize}
    \item The first in-depth investigation of the potential benefits of tactile information visualization;
    \item A  nuanced \amend{comparison} of tactile and visual information graphics perception and use;
    \item Identification of gaps in our understanding of the perception and use of tactile graphics and articulation of key research questions for the development of effective tactile information visualizations.
\end{itemize}

Our analysis makes it clear that a straightforward tactile translation of an effective visual graphic may not lead to an effective tactile representation and reinforces the need for a \textit{tactile-first} approach to the design of tactile visualizations~\cite{lundgard2021accessible,engel2021tactile}.

\section{Background} 
\label{sec:background}

In this section, we briefly examine the benefits of visual information graphics and then turn to research into accessible data visualization with a focus on tactile graphics.

\begin{figure*}
    \centering
    \includegraphics[width=0.67\linewidth]{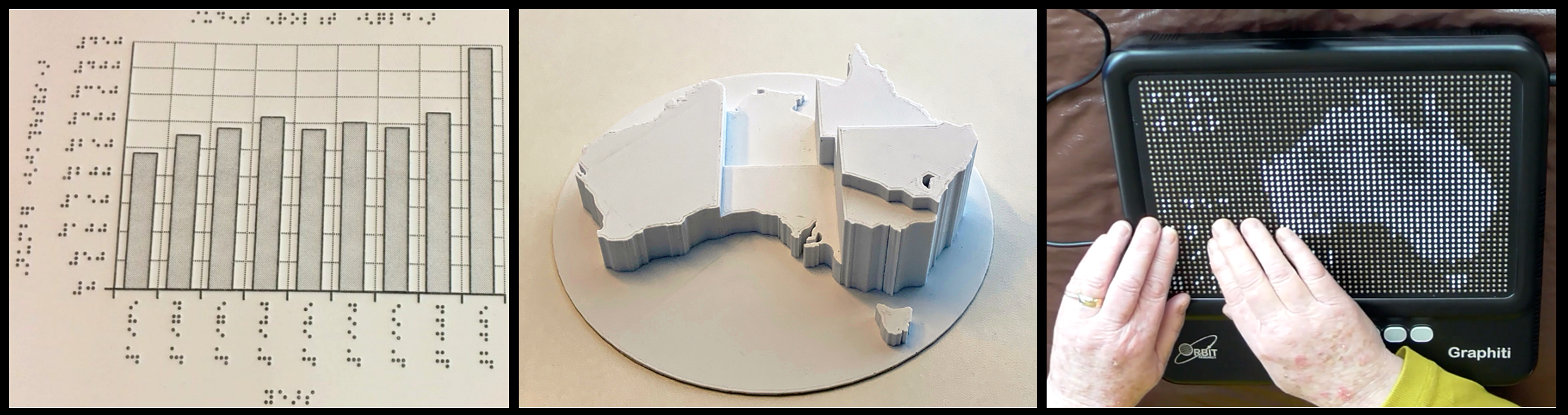}
    \caption{Tactile graphics are widely used to present information graphics to people who are blind or have low vision. Production methods include swell paper (left), 3D printing (middle), and refreshable tactile displays (right).}
    \label{fig:tactile-graphics}
    \vspace{-3mm}
\end{figure*}

\subsection{Benefits of Visualization}
There have been many suggested benefits of visual diagrams and visualizations. Some  originate in debates in the late twentieth century between AI and cognitive scientists about the differences between diagrammatic and sentential reasoning (e.g.,~\cite{glasgow1995diagrammatic}). These arguments were often mathematical or philosophical, and highlighted the congruence between visual representation and the data they represent~\cite{sloman1971interactions}, or the additional information made explicit in the visual representation~\cite{shimojima2015semantic}.

Other benefits relate to the properties of the visual system. 
Arguments of this kind are based on empirical experiments of visual perception~\cite{cleveland1984graphical,quadri2021survey,zeng2023review}, neurological data \cite{peck2013using,yoghourdjian2020scalability} and models of visual cognition  \cite{padilla2018decision,szafir2017modeling}. 

A third strand of research bridging the previous two considers the higher-level cognitive benefits of visual representations and visual reasoning. This is based on theoretical models of human reasoning~\cite{larkin1987diagram}, empirical experiments demonstrating, for instance, mental animation~\cite{hegarty1992mental} and ethnographic studies of diagram use~\cite{hutchins1995cognition}.

Streeb et al.~\cite{streeb2019visualize} analyzed these arguments to create a network of arguments for the use of visualization, where visualization was understood broadly to be any visual objects created for data analysis that use visual features beyond the type of sign and order of objects. They identified six classes of reasons: teamwork, cognition, interplay, design, semiotics, and basic. The potential benefits of well-designed visualizations and the underlying perceptual and cognitive reasons are also summarized in books on data visualization (e.g., Munzner~\cite{munzner2014visualization}, Ware~\cite{ware2021information}, and Marriott~\cite{marriott2024golden}).

\subsection{Accessible Data Visualization}

Recognizing the importance of making data visualization accessible to everyone~\cite{lee2020reaching}, there is a growing interest to improve access to data and data visualization for BLV people. Kim et al. conducted a comprehensive analysis of research papers published between 1999 and 2020, providing an overview of the approaches for accessible visualization design~\cite{kim2021accessible}. In addition to deriving a design space for accessible visualization, they identified key challenges and opportunities for future research.  \amend{Marriott et al. highlighted how different kinds of disabilities impact access to data visualization and the need for more research, including development of evidence-based guidelines for accessible data visualization}~\cite{marriott2021inclusive}. There also have been workshops (at Dagstuhl~\cite{lee2024inclusive} and VIS~\cite{hoque2024}) focusing on accessible data visualization aiming to establish a research agenda and build a community. A comprehensive review of accessible data visualization is beyond the scope of this paper. Instead, we briefly discuss a few key approaches. 

One common method for making graphical content accessible is through textual descriptions, often in the form of alternative (alt) text. 
There has been research aimed at improving textual descriptions specifically for visualizations. For example, Ault et al. developed and evaluated description guidelines for line charts~\cite{ault2002evaluation}. 
Gould et al. extended these principles to chart images, advocating for concise, clear, and data-focused descriptions structured in an overview-then-details manner~\cite{gould2008effective}. More recently, Lundgard and Satyanarayan proposed a four-level model for classifying the semantic content of natural language descriptions of visualizations~\cite{lundgard2021accessible}, and Jung et al. re-examined strategies for generating effective alt text for data visualizations by exploring the gap between existing guidelines and real-world implementation~\cite{jung2021communicating}. 

Sonification is another approach. This translates data into non-speech audio, allowing users to perceive patterns, trends, or relationships through sound~\cite{hermann2008taxonomy}. Flowers et al. examined the effectiveness of different auditory display techniques, demonstrating that simple auditory plots can effectively convey information about small data sets~\cite{flowers2005data}. This technique has been applied to interactive data analysis as an alternative to visual encoding for BLV people by mapping data attributes to auditory features, such as pitch, volume, and timbre~\cite{kramer2010sonification}. Many tools now exist for Cartesian-based charts, including SAS Graphic Accelerator (https://support.sas.com/software/products/graphics-accelerator), HighCharts (https://www.highcharts.com), and Apple’s Audio Graphs (https://tinyurl.com/audio-graphs). 

Researchers have explored the potential benefits of integrating speech and sonification. For example, Siu et al. explored audio data narratives, combining textual descriptions and sonifications 
for screen reader users~\cite{siu2022supporting}. 
Similarly, Holloway et al. compared standard sonification with infosonics, which have added spoken text information and labels~\cite{holloway2022infosonics}. 
Thompson et al. designed Chart Reader, a web-based accessibility engine that enables screen reader users to interactively understand data visualizations leveraging textual descriptions, sonification, and flexible navigation~\cite{thompson2023chart}. Sharif et al. developed VoxLens, \remove{an interactive JavaScript} a Chrome plug-in that leverages voice-based interaction and sonification to support online data visualizations for screen reader users~\cite{sharif2022voxlens}.

There has also been research into combining touch screen controlled sonification and speech that may also include \amend{vibratory} feedback~\cite{butler2021technology}. \amend{Such approaches can support direct spatial understanding, flexible interaction and layered information access~\cite{zhao2024tada}.} Goncu et al.~\cite{Goncu2011gravvitas} introduced GraVVITAS which provided a mix of audio and \amend{vibratory} feedback as the user explored the graphic on a touchscreen. Zhao et al. introduced iSonic, an interactive sonification system that enables BLV users to explore geospatial data through audio on a touchscreen~\cite{zhao2005isonic} \amend{and Zhao et al. developed a system for exploring node-link diagrams on a touchscreen using sonification and audio labels~\cite{zhao2024tada}.} 
Zhang et al.’s smartphone application, ChartA11y, integrates VoiceOver announcements and sonification  with \amend{vibratory} feedback when data points are contacted in scatterplots~\cite{zhang2024chart}.

\subsection{Tactile Graphics}

Raised line drawings known as tactile graphics~\cite{kim2021accessible} (Figure~\ref{fig:tactile-graphics}) are also widely used to provide accessible graphics to BLV people. Many BLV students encounter them at school, and they are often used in orientation and mobility training. \amend{Unlike pure linear audio or descriptions, tactile graphics provide a direct spatial representation and so
  accessibility guidelines recommend their use for presenting graphics such as maps and charts where spatial relationships are important}~\cite{BANA2022guidelines,DIAGRAM2013}. 

Nowadays, most tactile graphics are designed on the computer by professional transcribers and printed using specialised equipment and paper. In the simplest approach, a braille embosser punches raised dots into thick paper. Another common technique is to print the graphics onto swell paper. This contains microcapsules of alcohol that rise when heat is applied. Older techniques still in use include hand-drawing and collage, which are favoured in school settings and orientation and mobility training when precision is sacrificed for the speed of on-the-spot creation~\cite{Phutane2022}. 
Thermoform, a technique for making plastic vacuum-formed copies of hand-constructed tactile graphics, was once popular due to its low cost and ability to provide layered heights, but it is now losing favour due to the manual labour requirements.   

A recent development in tactile graphics has been the use of 3D printing. To date, the focus has been primarily on its use for conveying 3D educational graphics (in particular in STEM)~\cite{Koehler2017,Sun2017,Yazici2022} and 3D maps for use in orientation and mobility training~\cite{Nagassa2023,Wang2022}. However, it can also be used to print tactile information graphics~\cite{braier2014haptic,Gupta2019}.

However, none of these production technologies are well-suited to \amend{interactive} data visualization because of the time and cost to produce a single tactile graphic. Commercial refreshable tactile displays (RTDs) have recently emerged as an alternate tactile graphic presentation technology that is much better suited to such applications. Consisting of pin-arrays controlled by electro-mechanical actuators, RTDs have the ability to display a tactile graphic in a matter of seconds at essentially no cost. \amend{While RTDs remain expensive, it is likely that prices will continue to fall as more devices come on to the market. }

There has been relatively little research into accessible data visualization with RTDs. Reinders et al. investigated the combination of RTD with a speech-based conversational agent~\cite{reinders2024refreshable}. They conducted a Wizard-of-Oz study with 11 BLV participants, involving line charts, bar charts, and isarithmic maps. Many participants appreciated the tactile graphics for enabling independent analysis and the conversational agent for providing verification. 
Seo et al. co-designed and developed MAIDR~\cite{seo2024maidr}. This employs four modes—Tactile graphic (shown on a single line braille display), Text, Sonification, and Review—to provide an inclusive framework for interacting with bar charts, heat maps, box plots, and smooth-line-layered scatter plots.

Although transcription guidelines advocate  the use of tactile graphics when spatial layout is important to understanding the meaning of the graphic~\cite{BANA2022guidelines,DIAGRAM2013,RoundTable2022}, they do not explicitly explain the benefits of tactile graphics. In practice, many blind people anecdotally question the value of tactile graphics, which often prioritize visual concepts over representations that reflect how the world is experienced by touch. For example, a congenitally blind person may draw a circle to represent a tree, reflecting their experience of their arms encircling the tree’s trunk. 

Nonetheless, prior research comparing tactile visualizations with other modalities provides some initial confirmation for the value of tactile data visualizations. Most work has been in the area of mapping, with evidence that tactile maps provide superior understanding of spatial layout compared with physical exploration of the space~\cite{Caddeo2006} or a verbal description~\cite{Papadopoulos2018}. Loitsch \& Weber examined understanding of a UML sequence chart by seven BLV adults using either a text description or a refreshable tactile display with audio labels. They found that cognitive workload was similar for both representations, but some indication there may be fewer errors using a tactile representation~\cite{LoitschWeber2012}. Watanabe \& Mizukami found tactile representations of scatterplots enable quicker and more accurate identification of x-y relationships compared with tables of data~\cite{WatanabeMizukami2018}. Finally, Yang et al. found that tactile node link diagrams gave better performance for path following and cluster identification than a braille list of  node neighbors~\cite{Yang2020}.

However, this research has only examined particular types of tactile visualizations and not considered the underlying perceptual and cognitive reasons why tactile visualizations can be beneficial. A preliminary investigation by Goncu et al.~\cite{goncu2010tactile}  suggested potential benefits are: topological and geometric resemblance, computational off-loading/free rides, indexing, mental animation, more memorable and understandable symbols and possibly macro/micro view.  However, this work did not examine all benefits previously identified for visual graphics and only considered printed tactile graphics and so did not consider the benefits that interaction or animation might provide on an RTD.

Given the new focus by the visualization community on accessible visualizations and consequent interest in tactile graphics, it is timely to conduct a more thorough analysis of what benefits tactile visualizations might provide and the differences and similarities to those provided by visual information graphics.

\section{Our Framework}
\label{sec:approach}

In this section, we introduce a framework to investigate the benefits of visual and tactile information visualization and the differences between them. 
Our framework is based on Ware's simplified model of how we perceive and reason with data visualizations~\cite{ware2021information}. While this model was introduced for  visual information graphics, it applies equally to tactile visualizations. It has three components: 
\begin{itemize}
\item\textbf{Encoding:} Mapping data to a visual or tactile representation and displaying this in print on a computer monitor, RTD, etc.\footnote{While  encoding may have different meaning in vision science, here, we use the definition applied in visualization research. }
\item\textbf{Perception:} Low-level automatic processing of the image prior to conscious reasoning. 
\item\textbf{Cognition:} High-level conscious processing that integrates perceived information to address a task, resulting in insight or knowledge. 
\end{itemize}

We note that the distinction between perception and cognition is not clear cut but, for our purposes, we divide the two concepts chiefly between preconscious (i.e., the processes people perform automatically when looking at an image) and conscious reasoning (i.e., the processes that use the outputs of perception in combination with other information to form knowledge)~\cite{nes2023perception}.

This three-component model of how we perceive and reason with data visualizations provides a framework to investigate the benefits of visual and tactile information visualizations and the differences between them.

The first and most fundamental factor impacting visualization effectiveness is the \textit{encoding}. A good visual or tactile data encoding makes information beyond just the raw data values geometrically and visually or tactually explicit. For instance, 
in a network diagram,
the distance between nodes typically reflects how closely they are connected.

The second type of factors impacting visualization effectiveness are linked to \textit{perception}. In the case of a node-link diagram, for instance, visual perception will group nearby nodes into clusters and can quickly direct attention to red or flashing nodes.

The third type of factors impacting effectiveness are related to \textit{cognition}. This refers to the way a visualization supports conscious reasoning. \amend{It includes, for instance, the use of graphics to extend working memory and interactive manipulation of the visualization.}

\amend{Any particular benefit will generally involve aspects of encoding, perception, and cognition. However, linking the benefit to the primary encoding, perceptual, or cognitive factors that support it allows us to better understand which benefits are likely to hold for both tactile and visual information graphics versus for just one modality. }

\subsection{Visual Perception and Cognition}

Visual perception mechanisms extract the gist of the image, identify visual features such as edges and regions with uniform colour or texture and group these into a series of sensory percepts that enable higher levels of reasoning. These percepts are stored in memory and fused with other knowledge from memory to form insights. Visual perception and cognition uses both working memory and long-term memory.

Working memory is the short-term storage that allows people to rapidly recall and reason over what they have recently encountered, similar to a cache in a computer system.  Traditional visualizations likely primarily use visual working memory. Visual working memory holds the graphic elements of immediate or recent attention. These are typically a combination of external visual information (from iconic memory) and mental imagery. These elements are linked to information in long-term memory.

Working memory has very limited capacity, somewhere between~3 and~7 items \cite{brady2013probabilistic}. However, these items can be high-level \textit{chunks} that combine various pieces of information into a single unit, such as a statistical summary of a complex object or set of objects \cite{brady2016working}. In the case of visual working memory, these might be axes, trends, patterns, or other forms of statistical information rather than individual data points.

Other information is stored in long-term memory, from which knowledge relevant to the task at hand can be recalled. This information may comprise elements of the visualization, such as past states in an interactive visualization, or tangential information, such as theories, context, or expectations of the data.

\subsection{Haptic Perception and Cognition}
\label{sec:tactile_perception}

Like visual perception and cognition, haptic perception and cognition rely on working memory and long-term memory.
What we think of as a sense of touch or haptic perception is actually a combination of information from mechanoreceptors in the skin and subcutaneous tissues, mechanoreceptors in the joints and muscles, and free nerve endings that react to temperature and pain~\cite{HalataBaumann2008,HellerGentaz2014}. Skin receptors provide information about the object being touched while receptors in the joints and muscles inform how the body is positioned. 

While temperature can be perceived through passive touch, most other haptic properties require active touch for perception - deliberate movements such as force for hardness/softness, shearing movement for braille reading and textures, and enclosure for shape~\cite{Goodwin2008,KlatzkyLederman2007}. Localized properties such as temperature, curvature, and texture are perceived before global properties such as shape \amend{that are more salient in vision.}

At the most basic level, the haptic field is limited to the size of a single finger pad, with the pad of the index finger as the primary contact point for reading braille and tactile graphics. However, a well-trained touch reader will use multiple fingers at once, directing their attention to what is under one finger but using other fingers as a point of reference, to explore what else is nearby or move to the next area for reading.

\begin{figure}[t!]
    \centering
    \includegraphics[width=0.75\linewidth]{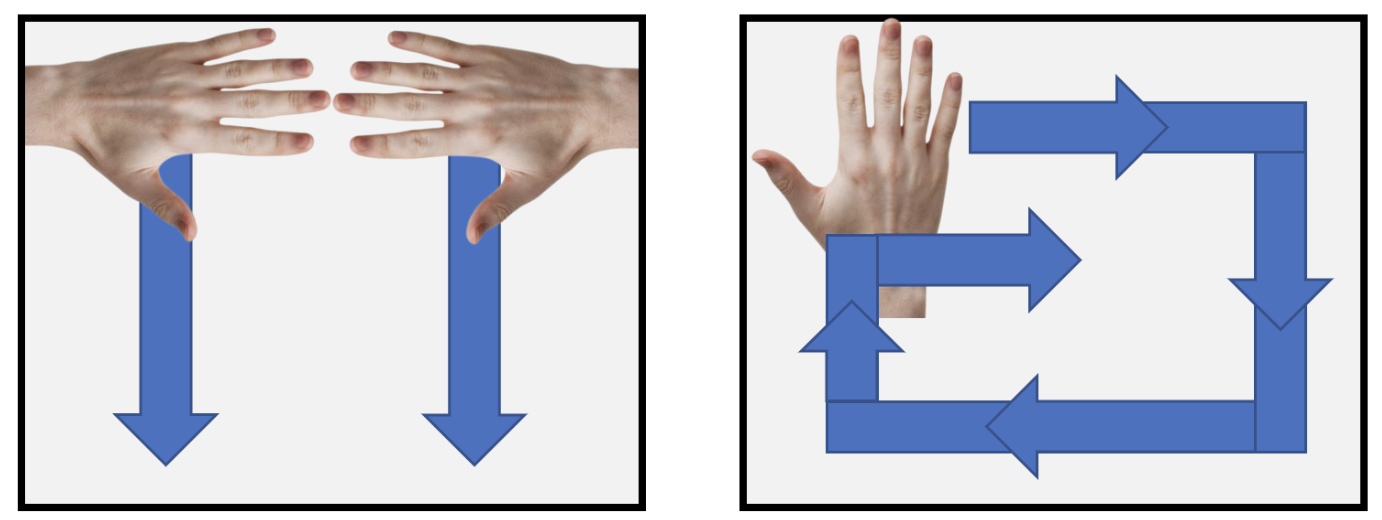}
    \caption{The waterfall technique (left) and the perimeter search (right) for systematic scanning of a tactile graphic to gain an overview of where features are located on the page.}
    \label{fig:overview-touch-reading}
    \vspace{-3mm}
\end{figure}

Overall, tactual reading can be best described as a bottom-up process, during which the reader must build up an image of a whole from the sum of its parts. Young readers are therefore taught strategies to gain a cursory impression of what is on the page before systematically exploring the graphic for more detail~\cite{Amick2022,Rosenblum2018,Withagen2010}. The most common overview methods are the \textit{waterfall technique} (Figure~\ref{fig:overview-touch-reading} left) -- placing both hands flat at the top of the page, with fingers meeting in the center, then moving the hands downwards to the bottom of the page~\cite{WeddeBishop2017} -- and the \textit{perimeter search} (Figure~\ref{fig:overview-touch-reading} right) -- scanning with one or two hands in a circle around the outside of the page before moving to the center~\cite{bardot2017identifying}. Both of these techniques utilize \textit{surface scanning} in which the whole hand is held palm-down on the page. This reduces tactile acuity but has the advantage of increasing the size of the haptic field.

Scanning provides an impression of the overall layout but does not give access to details, which must likewise be explored systematically to build up an overall picture. The most efficient process for exploring the details differs according to the type and arrangement of the graphic~\cite{Dulin2007,Zhao2021}. For example, for a graph, students will read the axis labels first before looking at the data trend and finally examining specific data points~\cite{Rosenblum2018}. Whereas, to understand a drawing, following lines is the most common strategy~\cite{Bara2013}. For other graphics such as maps, it may be more efficient to choose a reference point and explore other parts of the graphic in relation to this point~\cite{Zhao2021}.

Thus, the touch reader of a tactile graphic must direct their hand movements and attention to gather information while simultaneously holding this information in working memory to build up a picture of the graphic as a whole. 
It is now believed that tactile working memory has many similarities to visual working memory and that blind people, regardless of the age of onset of blindness, build functionally equivalent spatial images in working memory~\cite{cattaneo2008imagery,cattaneo2011blind}. In fact, both visual and tactile working memory may use a supramodal representation of space, which combines different sensory inputs into an abstract spatial representation with links to modality-specific information~\cite{cattaneo2011blind,schmidt2018brain}. However, this is not to say that spatial mental representations of blind and sighted people are identical. Indeed, research suggests that the use of touch favors egocentric representations while vision favors allocentric representations~\cite[p. 116]{cattaneo2011blind}. 

\subsection{Approach}

In the rest of the paper, we will examine arguments that have been made in the research literature for the effectiveness of visual information visualizations and consider to what extent these are likely to hold for tactile graphics. We use the framework introduced at the start of this section as a way of structuring this investigation.

The benefits and factors we discuss are drawn from Streeb et al.~\cite{streeb2019visualize}, Ware~\cite{ware2021information} and Munzner~\cite{munzner2014visualization} as well as our prior research\amend{~\cite{bae2025bridging,goncu2010tactile}}.
\amend{The choice of which benefits to consider} and whether they are likely to hold for tactile information graphics are, when possible, based on the research literature but also draw on our lived experience and professional expertise: two of the authors are congenitally blind and another author is an experienced tactile graphics producer. 

We acknowledge that our analysis is only a first step and that more research is required to test whether our conjectures and experiences generalize to the broader BLV community. \amend{Furthermore, as it has been guided by benefits identified for visual graphics, other benefits to tactile visualizations may have been missed. For instance, tactile representations may provide additional benefits due to physical embodiment.}

\section{Factors relating to Encoding}
\label{sec:encoding}

Encodings are at the heart of data visualization. These formalize the mapping between data and graphic representation. In the case of visual graphics, data elements and their attributes are mapped to graphical marks---points, lines, 2D areas, and 3D volumes---whose appearance is controlled by visual channels---position, color, shape, orientation, size (length, area or volume), texture, motion~\cite{munzner2014visualization,ware2021information}. 

Similarly, tactile encodings are at the heart of tactile visualizations. These employ the same graphical marks as visual encodings and have analogous tactile channels controlling their appearance, except that color is not available. However, not all channels are available on all tactile display media. For example, motion requires an RTD and 3D volumes require a format with variable heights.
Benefits primarily associated with encoding are:

\bstart{Geometric congruence~\cite{streeb2019visualize,goncu2010tactile,larkin1987diagram,kindlmann2014algebraic}:} The encoding can preserve (selected) geometric relationships between data elements. That is, the relative positions of data points can mirror the relative positions of the source data. For example, choropleth maps preserve the spatial location of data, while subway maps preserve the topology of the transport network.

\bstart{Animation~\cite{streeb2019visualize,jones2000animated,tversky2002animation}: }Animation allows dynamic processes and time series data to be shown with a congruent representation of time.

\bstart{Free-rides~\cite{goncu2010tactile,shimojima2015semantic}:} The mapping between data and encoding can make explicit relationships that are logical consequences of the original data. For example, because of the congruence between geometric containment and set containment, Euler diagrams 
show relationships between sets that are implicit in the data. 
An Euler diagram depicting that set $A$ is disjoint from set $C$ and that $B$ is a subset of $A$ necessarily shows that $B$ and $C$ are disjoint. As another example, placing connected nodes near one another in a node-link diagram ensures that geometric clustering correlates with network clustering.

\bstart{Indexing \& linking~\cite{goncu2010tactile,larkin1987diagram}:}  Visual and spatial indexing can be used for navigation and to link related elements. For example, data points can be found by using their spatial location as an index; proximity or indicator lines can link labels with graphical elements; color or texture can link elements with a legend;
and horizontal and vertical alignment link graphical marks to graph axes.

\vspace{1mm}\noindent
The above benefits are functions of design independent of how the visual or haptic systems work; regardless of whether the visual or haptic system can utilize these benefits, they are still inherent in the representation. Thus, they should be regarded as \textit{potential} benefits and the degree to which these benefits are realized depends on how well their implementation aligns with the capabilities 
of the visual and haptic perceptual systems.

For example, while animated transitions~\cite{archambault2016can,bederson1999does} and animated 2D vector fields~\cite{ware2016animated} have proven useful, in general, well-designed small multiples showing snapshots of a dynamic process often prove more effective than an animation of the process because of the demands animation makes on visual working memory and the limits of visual attention~\cite{tversky2002animation}. 

\section{Factors relating to Perception}
\label{sec:perception}

Prior research into visualization has identified several ways in which effective visualizations leverage visual perception. 
However, haptic and visual perception differ in ways that indicate that some of these benefits do not hold for the haptic perception of tactile graphics.

\bstart{Gist \& ensemble coding~\cite{bae2025bridging,padilla2018decision,ware2021information}:} When a viewer first looks at a visualization, automatic 
processing recognizes the \textit{gist} of an image in much less than a second, extracting information that likely includes the type of graphic and its main components as well as summary statistics about the distribution of visual features like size, shape, and color \cite{franconeri2021three,szafir2016four}. This processing triggers long-term memory associated with that graphic type, including what Pinker~\cite{pinker1990theory} calls the \textit{graphic schema}, the procedures and methods required to understand and interact with the graphic. As a result, the viewer can quickly understand the structure of a visual graphic and how to read it, and recognize salient high-level patterns.  

This is not the case for tactile graphics. As described in Section~\ref{sec:tactile_perception}, a touch reader must explore a visualization sequentially to build a mental image of the graphic to identify its type and main components. Unlike the automatic processes that interpret visual input, this requires conscious, intentional exploration and takes many seconds. 
 For this reason, guidelines recommend that tactile graphics come with a description of the graphic, including its type~\cite{BANA2022guidelines, DIAGRAM2013}, and touch readers are encouraged to read the caption or title before exploring the graphic~\cite{Rosenblum2018} as this allows the graphic schema to be triggered and to guide subsequent exploration and understanding of the graphic.

\bstart{Visual pop-out \&  search~\cite{streeb2019visualize,ware2021information,bae2025bridging}:} Visual information graphics can also direct attention to important information through both bottom-up feature-driven and top-down task-driven mechanisms.
The gist and ensemble codes produced when first looking at a visualization provide a sense of the distribution of visual features across the visualization's different spatial regions \cite{haberman2012ensemble}. As a result, readers have a preconscious mental model of the encoded data known as a \emph{feature map}.

This means that we rapidly identify objects that are perceptually distinct from their surroundings, such as red dots on a grey background or a moving object. These elements exhibit \emph{pop-out}: they immediately attract visual attention without the reader having to look for them. Pop-outs allow the visualization designer to guide attention to key components of the visualization. They also support fast parallel visual search for such salient data. This internal map of visual features can also drive more intentional search tasks by helping readers identify areas with high or low associations with a given visual feature~\cite{szafir2016four} (e.g.,  searching for a purple square in a cluster of mostly purple shapes before searching in a cluster of mostly red shapes). 

There has been some research into how haptic perception supports search. In a series of experiments, different types of target stimuli and distractors were placed beneath the fingertips. The only task demonstrating parallel search was distinguishing a bump from a flat surface~\cite{overvliet2007parallel}. 
Furthermore, the user must touch a graphic element to know if it is perceptually distinct from its background. Thus, the literature suggests that tactile pop-outs are much less useful than visual pop-outs. 
\amend{However, in the case of RTDs, we also know that vibrating or blinking pins are highly salient and sound can guide attention to their location~\cite{holloway2022animations}.}

\bstart{Perceptual organization~\cite{streeb2019visualize,ware2021information,bae2025bridging}:} One of the greatest benefits of visual perception in visualization is its ability to group visual elements into high-level structures. This process emerges from a mix of top-down visual attention and automatic pattern recognition. Perceptual psychologists have identified a number of visual Gestalt `laws’ of pattern recognition: proximity, similarity, connectedness, continuity, symmetry, closure and common region, and common fate. These play a critical role in recognizing clusters in a node-link diagram and trends and correlations in line graphs or scatter plots.

There has been much less investigation into haptic Gestalt laws and related mechanisms for structure estimation and pattern recognition. This is in part because psychologists initially believed that the sequential nature of haptic perception precluded them~\cite{gallace2011extent}. Recent research suggests that the laws of proximity~\cite{chang2007the,overvliet2013grouping}, continuation~\cite{chang2007gestalt}, similarity~\cite{chang2007the}, and symmetry~\cite{cattaneo2010symmetry} hold. However, many of these studies were conducted with sighted people and more research is required to understand haptic perceptual organization with touch readers.

\bstart{Attention~\cite{healey2011attention,goncu2010tactile,tufte1990envisioning}:} Visual attention refers to the specific, small region of a visualization that we actively process in detail (i.e., the encoded data we are directly looking at). 
Visual attention can be directed to the external environment or to visual working memory. We can attend to either a spatial region or an object. Visual attention is a little like a spotlight---we can choose to pay attention to a large region to see a trend, for example, or fixate on a smaller space to examine an object or small region in detail. 

There has been much less research into haptic attention, though it is believed to have similarities to visual attention~\cite{spence2007recent}. To some extent, haptic attention also behaves like a spotlight: touch readers can use the palms of their hands to attend to a larger area but with less resolution, use a fingertip to precisely examine a smaller area and even attend to small details by using a fingernail; however, this last strategy is usually avoided because it can be destructive to the tactile graphic. Nonetheless, the ability to zoom in and out is less fluid with touch reading, potentially making it difficult to directly perceive patterns or larger objects from the parts of the graphic being touched as  recognition of these patterns or objects relies on attending to images in tactile working memory. 

Another key difference is that visual attention focuses on one location at a time, with limited information about the surrounding data processed through peripheral vision. However, when touch reading, multiple fingers and hands can be used. For example, when comparing two time series in a line graph, it is common for touch readers to move an index finger from each hand along the different lines, attending to both lines and directly observing the similarities and differences. To perform the same process visually, a reader must scan their attention along each line individually and leverage working memory to try to compare the two lines. 

\bstart{Motion perception~\cite{ware2021information}:} The human visual system is highly sensitive to moving objects. When shown a rapid sequence of images with an object at different locations, we perceive this as a single moving object. We can use this ability to encode data \cite{munzner2014visualization} or trace key data points as a visualization changes \cite{huth2023studies}. Touch can also detect motion~\cite{KurokiNishida2021}, but requires the fingers to be in contact with the moving object. Thus, when showing animations on RTDs, successive frames must be displayed for long enough for the user to explore them and determine which objects have moved. This relies on working memory: there is not a perception of movement but rather a conscious identification of what has changed and of what movement caused this to happen. This may suggest that there is no point in animation and that it might be better to show all frames side-by-side \amend{in accord with Munzner's maxim that ``eyes beat memory''~\cite[p. 131]{munzner2014visualization}.} However, an advantage of showing the frames temporally is that touch readers can use exactly the same egocentric reference frame relative to the reader, making it easier for the reader to compare elements in adjacent frames~\cite{holloway2022animations}. \amend{More research is required to determine the relative benefits of each approach.}

\bstart{Channel effectiveness~\cite{ware2021information,munzner2014visualization}:} Many spatial visual channels have an associated perceptual ordering that supports a natural encoding of quantitative data. These include horizontal and vertical position, length, area and volume, curvature, and tilt. Some visual channels associated with color and texture also have an associated perceptual ordering, such as luminance, saturation, texture contrast, and scale. 
However, these channels are not equally effective at communicating precise information. For example, differences in length are more readily perceived than differences in volume or luminance \cite{cleveland1984graphical,heer2010crowdsourcing}. Graphical perception studies measure the effectiveness of different visual channels for a range of data analysis tasks (see Quadri \& Rosen \cite{quadri2021survey} for a survey). 

There has been little research into the effectiveness of different tactile channels for encoding quantitative data. A notable exception is work by Khalaila et al.~\cite{khalaila2024they} replicating prior studies into visual charts with BLV participants~\cite{cleveland1984graphical,heer2010crowdsourcing}. They compared the effectiveness of tactile bar charts, stacked bar charts,  pie charts and bubble charts. This found the tactile charts had the same effectiveness ranking as the corresponding visual charts. Xu et al. \cite{xu2023let} conducted a preliminary investigation into the relative effectiveness of different spatial channels and also found parallels between visual and tactile rankings with blindfolded users.

While some features of textures are likely to be perceptually ordered in touch, for instance, smooth-rough, sparse-dense, there has been little research into perceptually distinct dimensions for textures~\cite{picard2003perceptual}, and none that we are aware of into their relative effectiveness for displaying quantitative data to touch readers. Gupta et al. ~\cite{Gupta2019} compared the use of texture to height in tactile choropleth maps, finding a performance advantage for height.

Other visual channels, such as spatial region, hue, motion, texture and shape, are better suited to encoding categorical data.  Their relative effectiveness has been studied in a number of papers~\cite{tseng2024shape,demiralp2014learning,gleicher2013perception}.  Prior research has identified perceptually distinct tactile patterns for use in tactile graphics~\cite{Berla1972,CBA2003,JamesGill1975,Morris1971} but there have not been studies into the relative effectiveness of different tactile channels for distinguishing categorical data elements. It seems likely however, that tactually salient properties such as texture will be more effective than shape or size, which require tactile exploration and synthesis.

\bstart{Perceptual acuity:} 
The acuity of visual perception allows us to discern thousands or even millions of data points on a single screen \cite{fekete2002interactive}. 
Haptic perception has significantly lower resolution than visual perception. Two-point touch threshold tests on the finger pad yield distances of around 2-4 mm~\cite{LedermanKlatzky2009}. Accordingly, it is generally agreed that a minimum of 3 mm is required between lines and symbols for readability on a tactile graphic~\cite{Edman1992,NolanMorris1971} and point symbols should be a minimum of 5 mm in width~\cite{NolanMorris1971}. Preliminary studies suggest the need for increased minimum mark sizes and spacing between marks also applies for tactile visualizations \cite{engel2019user}. Our visual system can also discern a multitude of colors. 
However, the number of textures that can be distinguished tactually is limited, with guidelines recommending using no more than five different textures or fill patterns on a single diagram~\cite{BANA2022guidelines}. For these reasons, tactile graphics are often enlarged and simplified versions of their visual counterparts.

Furthermore, it is not possible to significantly increase the size of a tactile graphic to make up for its low resolution, as the size is restricted to the area that can be comfortably reached by the hands and by the limitations of the current presentation technologies. In particular, current RTDs have quite low resolution (e.g., the DotPad (https://pad.dotincorp.com/) 
 and Monarch (https://www.aph.org/meet-monarch/)
 provide 60$\times$40 and 96$\times$40 pins, respectively). 
This means that the information content of a typical tactile visualization is significantly less than that of a visualization displayed on a modern smartphone and much less than one shown on a standard monitor.

Issues of resolution and display size also impact the length of braille labels used in tactile visualizations. Braille must be a standard size and, unlike visual text, must be placed horizontally. This means that it is common to reduce the size of labels by using abbreviations and a key, increasing the cognitive effort required to read the label.

\section{Factors relating to Cognition}
\label{sec:cognition}

We now consider the ways in which tactile visualizations can support high-level conscious processing to foster insight or understanding. 

\bstart{Extend memory~\cite{streeb2019visualize,goncu2010tactile,ware2021information}:} One of the most obvious benefits of visualizations is that they extend the capacity of working memory.  In doing so, they amplify and scaffold cognitive processes by providing a meaningful external representation of the data for people to reason with~\cite{liu2008distributed}. This is as true for tactile visualizations as it is for visual information graphics, as tactile working memory has similar capacity limits to visual working memory. 

\bstart{\amend{Interaction~\cite{streeb2019visualize,munzner2014visualization}:}} \amend{Interaction allows the user to focus on relevant information at the appropriate level of detail by, for example, filtering, zooming and panning
and dynamically link data across visualizations using brushing. While traditional tactile graphics do not support interaction, RTDs do.}

\bstart{Communication \& collaboration~\cite{streeb2019visualize,hutchins1995cognition,liu2008distributed}:} Visual information graphics can be used to teach new representations of data, communicate insights to others, and support collaborative sensemaking~\cite{isenberg2011collaborative}. In theory, tactile visualizations offer similar potential benefits. However, practical and technical restrictions currently limit this. These 
include restricted access to tactile graphics and 
a limited ability to annotate tactile graphics, something that is useful in collaboration.

\bstart{Intuitiveness~\cite{marriott2024golden,ware2021information}:} 
Unconscious conceptual metaphors are a fundamental part of human reasoning. They allow us to leverage experience with the physical world to conceptualize abstract concepts like time or social relationships~\cite{lakoff2003metaphors}. Many visualizations leverage conceptual metaphors~\cite{marriott2024golden,pokojna2024language}: sociograms embody the social-relationship-is-a-physical connection metaphor,
blurriness leverages the uncertainty-is-a-fog metaphor, 
and line charts reflect the more-is-up or events-in-time-are-a-sequence-in-space conceptual metaphors. Conceptual metaphors can even influence the perception of magnitude (e.g., dark-is-more biases) or effectiveness of search (e.g., color-concept associations) \cite{schloss2024color}. Visualizations based on such metaphors may be easier to understand and seem more natural. 
It seems likely that visualizations relying on conceptual metaphors that are based on the experience of a blind person will also feel more natural and intuitive. However, these metaphors may vary drastically based on lived experiences of people with BLV \cite{holloway2023tacticons}.

\section{Synthesis}
\label{sec:synthesis}
In this section, we reflect on the implications \amend{for task performance} and tactile visualization design and then summarize the research gaps and questions that emerged from our analysis. 

\subsection{\amend{Task Performance}}
\label{sec:tasks}
The kinds of benefits exhibited by a visualization crucially depend upon the task the user wishes to  complete~\cite{albers2014task}. 
Here, we identify how differences in \amend{encoding, perceptual and cognitive factors 
impact the effectiveness of  tactile information graphics to support visualization tasks based on Munzner's ``Why'' framework~\cite{munzner2014visualization}. At a high-level, traditional visualizations and tactile information graphics support similar goals--discovery of insights, communication and  enjoyment. Where we see differences are in the lower-level tasks of \emph{search} and \emph{query}.  }

\bstart{Search:}~\amend{One of the most powerful benefits of visual information graphics is that they support data search and navigation.}
Effective search in visual information graphics relies on 
a mix of encoding, perceptual and \amend{cognitive} factors.
With respect to cognition,  \amend{\textbf{Interaction}} allows the user to filter, search or navigate to objects/regions of interest. 
For encoding, \textbf{Indexing \& linking }cues  support visual search.
Several perceptual factors also support search. Pop-out and feature maps capture the distribution and frequency of visual features across the entire visualization  (\textbf{Visual pop-out \& search}). Knowledge about the graphic’s structure contained in the graphic schema (\textbf{Gist \& ensemble coding}) and a hierarchical model of the graphic in working memory (\textbf{Perceptual organization}) give readers an intuition for where and how to scan the visualization for relevant information. Furthermore, the visual system remembers the position of objects recently held in attention in working memory, allowing immediate navigation back to these elements in the external graphic (\textbf{Attention})~\cite{ware2021information}.

\amend{Differences in perception mean that search is not as well supported by tactile visualizations. In particular, as discussed in Section~\ref{sec:perception}, tactile pop-outs are less useful than visual pop-outs. However, the other factors identified above--indexing \& linking, interaction, the graphic schema, a high-level model in working memory, and memory of the location of recently attended graphic elements--support tactual search.}

\bstart{\amend{Query:}}
\amend{Once the graphic elements  have been found, the user can query these elements to \emph{identify} information about a single element, \emph{compare} elements, or generate an \emph{overview} of multiple elements.}

Determining information about a single element in a traditional data visualization relies either on interactive interrogation, i.e., through a tooltip (\textbf{Interaction}), or navigation to where this information can be found, for example, through a label or on an axis (\textbf{Index \& linking}). Navigation is similar in both visual and tactile modalities \cite{engel2019user}, and interactive interrogation of tactile graphics, i.e., details on demand, has been provided using audio labels~\cite{butler2021technology}.

While encodings can explicitly enable known comparisons \cite{gleicher2011visual,gleicher2017considerations}, most comparison tasks with visual information graphics rely on serial attention to the elements a reader wishes to compare and comparing each external element to information in working memory. Tactile comparison can work similarly, with readers attending to individual elements and recalling them in sequence. However, unlike visual attention, which is restricted to a single fixation point at any given time, touch readers may  make use of the increased attentional space provided by using multiple fingers to attend to multiple points simultaneously, reducing demands on working memory (\textbf{Attention}). For instance, they can directly compare two elements by using a finger on each hand to explore the objects in parallel. The hands can also aid cognition in other ways. Khalaila et al.~\cite{khalaila2024they} observed touch readers using two fingers as a ruler or caliper to compare the length of bars in a bar chart or angle in a pie chart segment and splayed fingers to compare the area of bubble charts. Similarly, a study of tactile map reading strategies noted that touch readers used the span between two fingers or finger and thumb to measure distance~\cite{perkins2003real}.

Generating a visual overview of multiple elements can utilize both fast preconscious parallel processing, such as gist formation or Gestalt groupings (\textbf{Gist \& ensemble coding}, \textbf{Perceptual organization}), and slower conscious recognition of more nuanced patterns. Tactile summarization is slower and typically relies on working memory and more conscious processing of the graphic elements. 

\subsection{When are Tactile Graphics Likely to be Useful?}

The benefits of visual and tactile information visualizations due to encoding and cognitive factors are broadly similar. However, significant differences arise between the efficacy of tactile and visual information visualizations because of differences between haptic and visual perception. In particular, the  initial rapid preconscious overview of the graphic provided by sight (\textbf{Gist \& ensemble coding}) must be replaced by a slower conscious exploration of tactile graphics. Likewise, slower conscious exploration must be used to search for tactile elements (\textbf{Visual pop-out \& search}) and recognition of patterns and trends (\textbf{Perceptual organization}) relies on building an image in working memory. In short, touch readers must replace the use of fast, preconscious perception with slower, conscious spatial reasoning with high demands on working memory---in effect, replacing Type I (e.g., feature-driven) decision-making with Type II (e.g., task-driven) decision-making~\cite {padilla2018decision}.

Furthermore, tactile graphics cannot be as information-rich as visual graphics (\textbf{Perceptual acuity}). This means that tactile graphics such as bar charts or node-link diagrams are only suitable for small data sets. 
RTDs offer a way around this as they (potentially) support interactive exploration of larger data sets (\textbf{Interaction}). Indeed, because of the limits of tactile acuity,  interaction is potentially even more important for touch readers than it is for sighted readers.

\amend{These differences and our analysis of task performance suggest that tactile graphics will not always provide similar benefits to visual graphics and that even when 
a visual graphic is beneficial, a touch reader may be better off using a simple data table due to the high cognitive overhead of reading a more complex tactile visualization. }

\amend{The use of dashboards and control panels for real-time visual monitoring provides a good example of how differences in visual and haptic perception impact  effectiveness.    Visual monitoring crucially relies on preconscious processing to direct visual attention to changes, which can be designed to pop-out or attract attention (\textbf{Visual pop-out \& search}) to avoid change blindness~\cite{stehle2020real}. While research has explored how screenreaders might support monitoring dashboards for BLV \cite{srinivasan2023azimuth}, the reliance on attentional salience in detecting changes means that tactile visualizations in isolation are likely to be ill-suited to this task. }

Despite these differences, as we have seen, tactile visualizations can be beneficial.  Our analysis allows us to understand why.
As an example, consider the study by Yang et al.~\cite{Yang2020}, which compared four tactile representations of (social) networks: an organic node-link diagram, a node-link diagram laid out on a grid, an adjacency matrix, and a list of neighbors for each node.
The 11 blind participants more accurately determined the number of clusters in the networks with the two node-link representations than with the list or matrix. This is because, in the encoding, spatial clustering reflected social clustering (\textbf{Free-rides}) and haptic perception built a high-level representation of the network based on spatial proximity (\textbf{Perceptual organization}).
The most difficult task for participants was to find the shortest path between two nodes. They were faster and more accurate with the node-link representations, using the links and spatial position of the nodes to guide their search for a path between them (\textbf{Indexing \& linking}).
None of the participants  had seen any of these representations before but rated the node-link diagrams as more understandable, imaginable, and intuitive than the other two representations. This may be explained by the fact that the node-link representations reflect the social-relationship-is-a-physical-connection metaphor (\textbf{\amend{Intuitiveness}}).

These benefits are similar to those previously identified for visual node-link diagrams~\cite{bae2025bridging}. Yang et al.'s study, however, also suggested some differences. When participants were asked to estimate the number of links in the network, they consistently underestimated the number of links when using node-link representations. This is in line with the difficulty of building an initial overview of a tactile graphic, as participants could not automatically gain a sense of the edge density at a glance 
(\textbf{Gist \& ensemble coding}). The study also found that many participants did not realize that the adjacency matrix was diagonally symmetric. While this is visually obvious, it was not tactually perceived, probably because the mental image was not detailed enough to reveal the symmetry (\textbf{Perceptual organization}). It was also noted how when asked to identify the number of neighbors in common between two nodes, participants used two fingers, running these in parallel along the corresponding rows or columns in the matrix representation (\textbf{Attention}). Reasoning about explanatory factors for these differences helps guide development of guidelines for effective tactile data representation.

\subsection{Gaps and Research Agenda}

Our exploration has revealed significant research gaps and opportunities. It has also revealed that much of what we know about tactile perception and cognition is based on studies with sighted people. It is unclear to what extent the findings from these studies generalize to BLV people and how the age of advent of blindness affects the findings. 

\bstart{Perception of tactile graphics:} In comparison to the perception of visual graphics, we know very little about the low-level haptic processing of tactile graphics. When and to what extent do Gestalt laws describe perceptual organization in haptic perception? 
What is the role of working memory in the recognition of larger patterns and objects? Which haptic features are most salient---for instance, vibration or height---and so are well-suited to guide attention? 
Which tactile channels are most effective at showing quantitative and categorical information? All of these are key questions that need to be answered before we can have a scientific basis for designing tactile visualizations.

\bstart{Touch reading strategies:} We also have little understanding of the comparative benefits of different touch reading strategies for interpreting tactile information graphics. What is the best strategy to build an initial overview? We have seen that touch readers can use their hands to reduce the cognitive load of comparison, either by exploring the items in parallel with two hands or by using a finger span as a memory aid. How effective are these strategies and how do we design tactile visualizations that take advantage of them to reduce cognitive load?

\bstart{Use of audio to complement tactile graphics:} While we have focused on stand-alone tactile graphics, there has been considerable research into the use of audio labels with tactile graphics~\cite{butler2021technology}. This reduces the need for space-filling braille labels and keys, and supports the use of tactile graphics by non-braille readers. Our investigation suggests that the combination of audio and tactile may have other benefits. Can we use sonification to assist in creating an initial overview of the graphic? Can we use speech and sonification to provide contextual information when the graphic changes on an RTD? Can we use spatial audio to support search? Can we use audio to increase information richness and reduce the impact of low tactile acuity?

\bstart{Research into RTDs:} Our research highlights that interactive exploration is essential for all but the smallest data sets. However, 
there has been little research into the use of RTDs for interactive data visualization. What are the best ways to support standard visualization interaction techniques like panning and zooming, filtering, brushing and linking? Can RTDs actually support real-world exploration of large data sets by BLV analysts?

There also needs to be more research into the design of RTDs to overcome  current limitations. Compared with tactile graphics produced through other methods, current RTDs have very low resolution. This impacts readability as lines are not smooth but rather a series of dots, and also limits the ability to show textured patterns. Furthermore, the pins on devices such as the DotPad or Monarch raise to a uniform height, preventing the use of height to convey information such as importance, depth, or categories.

\bstart{Collaboration:} Sense making is often collaborative~\cite{liu2008distributed,isenberg2011collaborative}. RTDs open the possibility of blind and sighted analysts working together with shared information graphics that are presented to each analyst using the appropriate modality. This raises many questions. How do we support automatic translation between these two modalities? How do we support annotation and shared understanding?   

\bstart{Scientific visualization:} We have focused on information visualization, but what about scientific visualizations in which 3D spatial data is visualized using techniques such as volume rendering? To the best of our knowledge, tactile graphics have rarely been used for this purpose. One reason may be the difficulty that many blind people have in understanding tactile drawings of 3D objects~\cite{HellerGentaz2014,picard2012identifying}. Visual conventions such as perspective or a single viewpoint make little sense to people who have never experienced sight. Touch readers must be taught these conventions, and even then, understanding a drawing of a 3D object requires attention, unlike the perception of visual depth cues by sighted viewers. \amend{Tactile drawing conventions not based on visual conventions can help but understanding these still remains cognitively demanding~\cite{panotopoulou2020tactile}.} Thus, it is an open question as to whether tactile graphics are suited to visualization of 3D spatial data or if other technologies, such as virtual reality or 3D models, may be a better alternative.

\subsection{Tactile-First Design}

Tactile transcription guidelines tend to advocate for replicating the visual graphic rather than redesigning for touch access.
This is partly due to an assumption that one of the roles of tactile graphics is to teach BLV students about visual graphics. However, the differences we have identified between visual and haptic perception mean that a straightforward tactile translation of an effective visual graphic may not lead to an effective tactile representation. If our goal is to support data exploration by touch readers, we need to design tactile information graphics with a \textit{tactile-first} mindset~\cite{lundgard2021accessible,engel2021tactile}.

\amend{Such a mindset can begin with broad general recommendations for the design of tactile graphics that are already widely agreed upon~\cite{BANA2022guidelines,Edman1992,RoundTable2022}, namely:} 
\begin{enumerate}
\itemsep -.2em
    \item[R1] \amend{Use a simple design (\textbf{Perceptual acuity}). }
    \item[R2] \amend{Provide an overview before the graphic. Ideally, this should include a description of the type of graphic, its subject matter and the recommended reading direction (\textbf{Gist \& ensemble coding}).}
\end{enumerate}

\amend{Based on our analysis, we can also suggest preliminary further recommendations for touch-first design. These may entail a complete re-design rather than a mere simplification of the visual graphic:}
\begin{enumerate}
\itemsep -.2em
\item[R3] \amend{Provide a predictable structure to support building a mental model and navigation (\textbf{Gist \& ensemble coding}, \textbf{Indexing \& linking}).} 
\item[R4] \amend{Locate information to reduce navigation requirements, even if this means introducing redundancy 
(\textbf{Indexing \& linking}).} 
\item[R5] \amend{Use tactile saliency to draw attention to important features, e.g., using height~\cite{Gupta2019} or flashing pins~\cite{holloway2022animations} (\textbf{Visual pop-out \& search}).}
\item [R6]\amend{Design for two-handed tactile exploration and comparison (\textbf{Attention}).}
\item [R7] \amend{Orient the diagram to maximise space for horizontal braille labels (\textbf{Perceptual acuity}).}
\item [R8] \amend{Consider using interaction to  support user-controlled data selection and details on demand (\textbf{Interaction}, \textbf{Perceptual acuity})}. 
\end{enumerate}

\begin{figure}
    \centering
    \includegraphics[width=\linewidth]{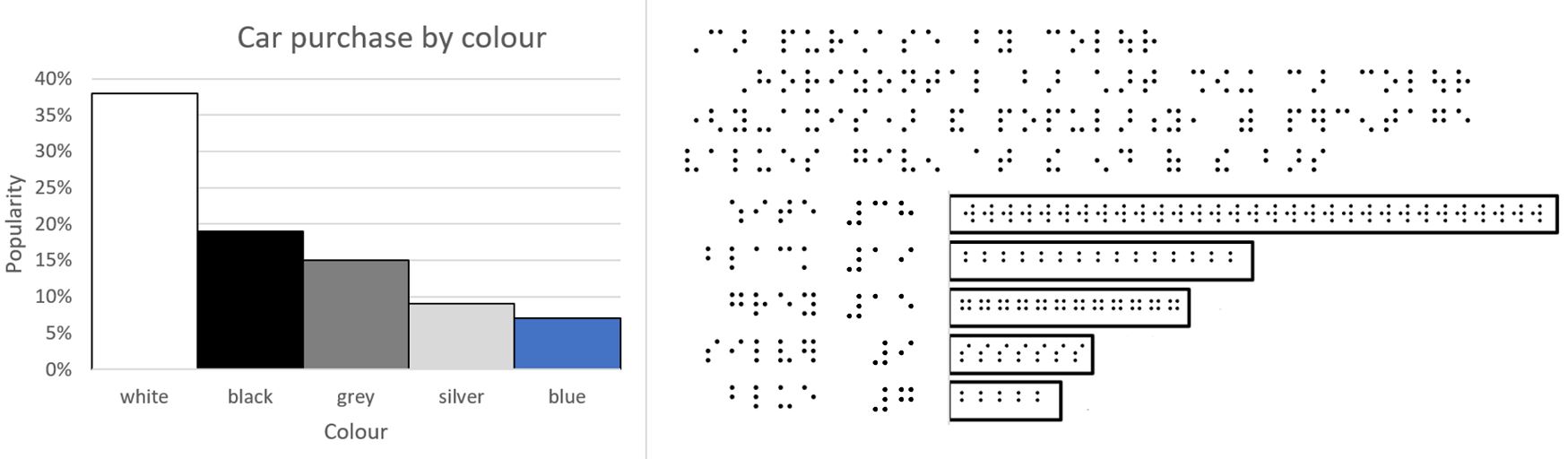}
    \caption{A histogram redesigned in tactile form to better facilitate touch access and understanding. }
    \label{fig:bar_chart}
    \vspace{-3mm}
\end{figure}

Figure~\ref{fig:bar_chart} provides an example. The novel tactile chart on the right represents the frequency data shown in the  histogram on the left. It includes a braille description of the type of chart and reading direction \amend{(\textbf{R2})}. The format has been flipped from columns to bars to allow more space for braille bar labels \amend{(\textbf{R7})}. The value labels are placed next to the bar labels to remove the need to move the finger to the axes \amend{(\textbf{R4})}. Fill colours for quick identification of the bars are replaced with a braille fill pattern giving the label's first letter, for example w for white \amend{(\textbf{R4})}. While testing is required to determine if this is, in fact, an effective tactile visualization, it illustrates how our analysis can inform the development of truly tactile-first designs.

As another example, small multiples are unlikely to be very useful for touch readers because of the limited amount of information that can be shown in a tactile graphic (\textbf{Perceptual acuity}). 
A better approach will be to allow the touch reader to interactively switch between the visualizations (\amend{\textbf{R8}}) or show a pair of multiples side-by-side to better facilitate two-handed comparison (\amend{\textbf{R6}}).

The design of an effective tactile data visualization also crucially depends upon the user. Late onset of vision loss can limit tactile literacy and acuity but mean that the person has had exposure to, and developed an understanding of, key visual representations.
Congenital blindness may mean that data visualizations such as node-link diagrams and scatter plots are unfamiliar. 
Tactile graphics design therefore should consider the age and likely skills of the intended reader~\cite{RoundTable2022,Wright2008}.

\section{Conclusion}
\label{sec:conclusion}

We have presented the first systematic investigation of the potential benefits of tactile information graphics. We first identified known benefits of visual information graphics and considered to what extent these hold for tactile graphics. We structured the discussion into benefits associated with \amend{encoding}, perception, and cognition. This has provided a nuanced understanding of the similarities and differences between tactile and visual information graphics in terms of perception and use.
Our investigation provides a theoretical basis for further investigation into tactile information graphics and the design of effective tactile representations. We have also identified key gaps in the current understanding of the perception and use of tactile graphics that we must address to develop a scientific basis for the tactile-first design of information graphics.

\acknowledgments{%
	The first author thanks Lace Padilla and Arvind Satyanarayan for early feedback on these ideas.
  We gratefully acknowledge the Australian Research Council’s Discovery Projects funding scheme (DP220101221).
  This work was supported in part by the Institute of Information and Communications Technology Planning and Evaluation (IITP) Grant funded by the Korean Government (MSIT), Artificial Intelligence Graduate School Program, Yonsei University, under Grant RS-2020-II201361, and the US National Science Foundation IIS \#2046725. 
}

\bibliographystyle{abbrv-doi-hyperref}
\bibliography{references,references2}

\end{document}